\documentclass[12pt]{article}
\usepackage{graphicx}
\def \b{{\cal B}} 
\def \bea{\begin{eqnarray}}
\def \beq{\begin{equation}}
\def \eea{\end{eqnarray}}
\def \eeq{\end{equation}}

\def \s{\sqrt{2}} 
\def \st{\sqrt{3}} 
\def \sx{\sqrt{6}}

\begin{document} 
\rightline{EFI 09-03} 
\rightline{TECHNION-PH-2009-03}
\rightline{arXiv:0902.1363} 
\rightline{February 2009} 
\bigskip
\centerline{\bf $\omega-\phi$ MIXING AND WEAK ANNIHILATION IN $D_s$ DECAYS} 
\bigskip 
 
\centerline{Michael Gronau\footnote{gronau@physics.technion.ac.il}} 
\centerline{\it Physics Department, Technion -- Israel Institute of Technology} 
\centerline{\it 32000 Haifa, Israel} 
\medskip 
 
\centerline{Jonathan L. Rosner\footnote{rosner@hep.uchicago.edu}} 
\centerline{\it Enrico Fermi Institute and Department of Physics} 
\centerline{\it University of Chicago, 5640 S. Ellis Avenue, Chicago, IL 60637} 
 
\begin{quote} 
The mixing between nonstrange and strange quark wavefunctions in the 
$\omega$ and $\phi$ mesons leads to a small predicted branching ratio $\b(D_s^+ 
\to \omega e^+ \nu_e) = {\cal O}(10^{-4})(\delta/3.34^\circ)^2$, 
where $\delta$ is the mixing angle.  The value $\delta = -3.34^\circ$ 
is obtained in a mass-independent analysis, while a mass-dependent analysis 
gives $\delta = -0.45^\circ$ at $m(\omega)$ and $-4.64^\circ$ at $m(\phi)$. 
Measurement of this branching ratio thus can tell whether the decay is 
dominated by $\phi$--$\omega$ mixing, or additional nonperturbative processes 
commonly known as ``weak annihilation'' (WA) contribute.  The role of WA
in the decay $D_s^+ \to \omega \pi^+$ and its possible use in estimating WA
effects in $D_s^+ \to \omega e^+ \nu_e$ are also discussed.
Assuming that the dynamics of WA in $D_s^+ \to \omega \pi^+$ 
is similar in $D_s^+ \to \omega e^+ \nu_e$ 
we estimate $\b(D_s^+ \to \omega e^+ \nu_e) = (1.3 \pm 0.5)\times 10^{-3}$.
\end{quote} 
 
\leftline{PACS numbers: 13.20.Fc, 13.25.Ft, 14.40.Lb, 12.39.Hg} 
\bigskip 
 
\centerline{\bf I.  INTRODUCTION} 
\bigskip 
 
The CLEO Collaboration has completed a study of $e^+ e^-$ production of 
charmed mesons near threshold, including a sample of about 600 pb$^{-1}$ 
at $\sqrt{s} = 4.17$ GeV \cite{Alexander:2009ux} where $D^+_sD^{*-}_s +
D^-_sD^{*+}_s$ pairs are produced with a cross section approaching 1 nb.  This
permits the study of rare $D_s$ meson decays. 
 
A prominent semileptonic decay of $D_s$ mesons is the process $D_s^+ \to 
\phi e^+ \nu_e$, with branching ratio \cite{Amsler:2008zzb} 
\beq
\b(D^+_s \to \phi e^+ \nu_e) = (2.36 \pm 0.26)\%~.
\eeq
At the quark level, this is represented by the process 
$c \to s e^+ \nu_e$, with the final $s$ and the spectator $\bar s$ forming 
a $\phi$.  One anticipated contribution to the process $D_s^+ \to \omega e^+ 
\nu_e$, on the other hand, is expected to involve the small $s \bar s$ 
admixture in the $\omega$ wave function, and hence to be highly suppressed by 
the Okubo-Iizuka-Zweig (OZI) \cite{Okubo:1963fa} rule.  An additional process, 
commonly known as ``weak annihilation'' (WA) \cite{WA}, would involve 
nonperturbative pre-radiation of an $\omega$ meson by the $c \bar s$ system, 
followed by the annihilation process $c \bar s \to e^+ \nu_e$.  A similar WA
process can account for at most a few percent of $B$ meson semileptonic decays
to charmless final states \cite{CLEOWA,Aubert:2007tw}.  The effects of WA in
semileptonic $D_s$ decays are expected to be considerably larger than in
semileptonic charmless $B$ decays.  We shall estimate their contribution to
$D_s^+ \to \omega e^+ \nu_e$ and discuss their characteristic kinematic
signatures.

One should distinguish between two types of WA often discussed in the
literature.  In $D_s$ decays, if the $c$ and $\bar s$ annihilate weakly into
$u \bar d$, and the $u \bar d$ state then materializes into a
non-strange final states such as $\pi^+ \pi^+ \pi^-$, there is, in principle,
no OZI suppression, although helicity conservation arguments for the light
$u$ and $\bar d$ quarks lead one to expect a suppression of the amplitude
unless at least two gluons also pass from the initial to the final state
\cite{Lipkin:1989vp}.  Such WA processes (when exchange amplitudes, which
are also in principle subject to helicity suppression, are included) are
likely to be a major source of charmed particle lifetime differences.
On the other hand, we are considering a form of WA which involves
OZI-suppressed nonperturbative pre-radiation of an isoscalar system such
as an $\omega$ meson, e.g., in $D_s \to \omega (D_s^*)_{\rm virtual} \to
\omega \ell \nu$.  It is this type of WA whose contribution to $B$ charmless
semileptonic decays can affect the extraction of $|V_{ub}|$ from such
processes \cite{WA}.

A recent discussion of the effects of $\omega$--$\phi$ mixing in $B$ meson 
decays may be found in Ref.\ \cite{Gronau:2008kk}.  The physical states may be 
represented in terms of ideally mixed states $\omega^I \equiv (u \bar u + d 
\bar d)/\s$, $\phi^I \equiv s \bar s$ by 
\beq\label{mixing} 
\left( \begin{array}{c} \omega \\ \phi \end{array} \right) = 
\left( \begin{array}{c c} 
\cos \delta & \sin \delta \\ - \sin \delta & \cos \delta 
\end{array} \right) 
\left( \begin{array}{c} \omega^I \\ \phi^I \end{array} \right) 
\eeq 
 
In one mass-independent analysis \cite{Benayoun:1999fv} a mixing angle 
$\delta = - (3.34 \pm 0.17)^\circ$ was obtained, while allowing for 
energy dependence \cite{Benayoun:2007cu} one finds $\delta$ varying from 
$-0.45^\circ$ at $m(\omega)$ to $-4.64^\circ$ at $m(\phi)$.  A measurement 
of $\b(D_s^+ \to \omega e^+ \nu_e)$ can help distinguish between these 
predictions and uncover any new effects beyond those associated with 
$\phi$--$\omega$ mixing.  In this article we predict the relation between 
$\b(D_s^+ \to \omega e^+ \nu_e)$ and the mixing angle, and note how further 
data on $D_s^+ \to \phi e^+ \nu_e$ can sharpen the prediction, potentially
providing also information on a contribution of weak annihilation.  We 
also discuss the role of weak annihilation in the hadronic two-body decay
$D_s^+ \to \omega \pi^+$.

The treatment of $D_s$ decays benefits from a heavy-quark symmetry framework
described in Section II.  The approach is applied here to the semileptonic
decays $D_s^+ \to (\omega,\phi,\eta,\eta') \ell^+ \nu_\ell$, and to the two-body
hadronic decays $D_s^+ \to (\omega,\phi,\eta,\eta') \pi^+$ in Section III.  We
discuss a possible connection between WA in $D_s^+ \to \omega \pi^+$ and
$D_s^+ \to \omega e^+ \nu_e$ in Section IV, and conclude in Section V.
\bigskip 

\centerline{\bf II.  SEMILEPTONIC $D_s$ DECAYS} 
\bigskip
\noindent
{\bf A. Kinematic and form factor effects} 

\medskip
We begin by comparing kinematic and form factor effects on the decays $D_s \to
(\omega_s,\phi) \ell \nu$, where $\omega_s$ is a ficitious particle with the
mass of $\omega$ and the pure-strange-quark content of an ideally mixed
$\phi_I$.
The phase space for decay of a particle of mass $M$ to three final-state 
particles, one of which has mass $m$ and the other two of which are massless, 
is reduced with respect to that for three massless final particles by a factor 
$g(x) \equiv 1 - x^2 + 2 x \ln x$, where $x \equiv (m/M)^2$.  Applying this
to the decays $D_s \to (\omega,\phi) \ell \nu$, we find $x = (0.158,0.268)$
and $g(x) = (0.392,0.222)$ for $(\omega,\phi)$.  The ratio of these two
values is 1.76.  This kinematic factor is the one appropriate for a flat Dalitz
plot.  For the decay of a fermion of mass $M$ to another of mass $m$ and two
massless fermions with $(V-A) \times (V-A)$ coupling, as in $\tau^- \to \mu^-
\nu_\tau \bar \mu_\nu$, the appropriate function would instead be
$f(x) = 1 - 8x + 8 x^3 - x^4 - 12 x^2 \ln x$, equal to $(0.3195,0.1395)$ for 
$(\omega,\phi)$.  The ratio of these last two values is 2.3. 
 
Form factors can affect the ratio $R_s \equiv \Gamma(D_s \to \omega_s \ell
\nu)/\Gamma(D_s \to \phi_I \ell \nu))$.  In the heavy-quark formalism of Refs.\
\cite{HQS}, as applied in Ref.\  \cite{Rosner:1990xx}, a
single form factor governs all the helicity amplitudes for the decays of a
pseudoscalar meson to a vector or pseudoscalar meson and a lepton pair.  We
employ this formalism primarily to illustrate the possible variations from the
flat-Dalitz-plot value of $R_s=1.76$.  We find a range $1.2 \le R_s \le 2.4$
in various applications of the symmetry, and shall consider these rather
conservative bounds in estimating the rate for $D_s \to \omega \ell \nu$ due
to $\omega$--$\phi$ mixing. 
 
To the extent that the strange quark's effective mass in $D_s \to \phi \ell
\nu$ can be regarded as 0.5 GeV/$c^2$ (its ``constituent-quark'' mass), the
heavy-quark limit discussed in Ref.\ \cite{Rosner:1990xx} begins to have some
validity.  Its limitations for the related processes $D \to (\overline{K}^*,
\overline{K}) \ell \nu$ were discussed extensively in Ref.\
\cite{Amundson:1992xp}.  Primary among these limitations is the importance of
$1/m_s$ corrections in reducing the predicted branching ratio for
$D \to \overline{K}^* \ell \nu$ by about a factor of two while affecting
the predicted branching ratio for $D \to \overline{K} \ell \nu$ much less.
We shall see that similar effects are called for when comparing $D_s \to \phi
\ell \nu$ with $D_s \to (\eta,\eta') \ell \nu$.
A form factor parameter which describes $D_s \to (\eta,\eta') \ell \nu$
adequately will be seen to predict a branching ratio for $D_s \to \phi
\ell \nu$ about a factor of 2 above experiment.  As $D_s$ semileptonic decays
are related to $D$ semileptonic decays by replacement of a nonstrange by a
strange spectator quark, we expect the pattern of $1/m_s$ corrections in
the former to be the same as in the latter, for which a satisfactory
description was obtained \cite{Amundson:1992xp}.

The universal form factor $\xi(w^2)$ is a function of the invariant square 
of the universal velocity transfer $w = v - v'$, with $v = p_{D_s}/M_{D_s}$ 
and $v' = p_V/m_V$.  In terms of the invariant square $q^2 = m^2_{e \nu}$ of 
the 4-momentum transfered to the lepton pair, one has 
\beq \label{eqn:wsq}
w^2 = \frac{q^2 - q^2_{\rm max}}{M_{D_s} m_V} = \frac{q^2 - (M_{D_s} - m_V)^2} 
{M_{D_s} m_V}~. 
\eeq 
The form factor $\xi(w^2)$ is normalized to unity at $w^2 = 0$ aside from a 
QCD enhancement factor $E$ \cite{QCDcorr}: 
\beq \label{eqn:xi}
\xi(w^2) = \frac{E}{1 - w^2/w_0^2}~,~~ E \equiv 
\left[ \frac{\alpha_s(M^2_{D_s})}{\alpha_s(m^2_{V})} 
\right]^{-6/(33 - 2 n_f)}~. 
\eeq 
We take the number of flavors $n_f$ equal to three.  For our purposes it is 
sufficient to estimate $\alpha_s(M^2_{D_s}) = 0.3$, $\alpha_s(m^2_{V}) = 0.4$, 
so the QCD enhancement factor is $E = (3/4)^{-2/9} = 1.066$.  

The differential 
decay rates with respect to the dimensionless parameter $y = q^2/M^2_{D_s}$ 
for pseudoscalar mesons $P$ and for transversely and longitudinally polarized
vector mesons $V_{T,L}$ are then \cite{Rosner:1990xx} 
\beq 
\frac{d \Gamma_p}{d y} = \frac{\Gamma_0 \lambda^{1/2}(1,\zeta,y)f_p(y)} 
{(1 - w^2/w_0^2)^2}~, 
\eeq 
\beq 
\Gamma_0 \equiv \frac{(G_F V_{cs} E)^2 M_{D_s}^5}{192 \pi^3} = 7.28 \times 
10^{-13}~{\rm GeV} 
\eeq 
for $V_{cs} = 0.974$ and $M_{D_s} = 1968.5$ MeV/$c^2$.  Here $\lambda(a,b,c) 
\equiv a^2 + b^2 + c^2 - 2ab - 2ac - 2bc$, $\zeta \equiv m_V^2/M^2_{D_s}$, 
while 
\beq 
f_p(y) \equiv \left\{ \begin{array}{c}
(1 + \sqrt{\zeta})^2 \lambda(1,\zeta,y)/4\sqrt{\zeta}~~~~~~~~~~~~~ p=P e \nu~, \cr 
y[(1 + \sqrt{\zeta})^2 - y](1 + \zeta - y)/\sqrt{\zeta}~~~~~p=V_T e \nu~, \cr 
(1 - \sqrt{\zeta})^2[(1 + \sqrt{\zeta})^2 - y]^2/4\sqrt{\zeta}~~~~~p=V_L e \nu~. 
\end{array} 
\right. 
\eeq 
For $\tau_{D_s} = 0.500\pm0.007$ ps \cite{Amsler:2008zzb}, the corresponding
differential branching ratios with respect to $y$ are then $dB_p/dy = 0.55 
\lambda^{1/2}(1,\zeta,y)f_p(y)/(1 - w^2/w_0^2)^2$.

In Ref.\ \cite{Amundson:1992xp}, a parametrization for the universal
monopole form factor was adopted with $w_0 = \sqrt{2}/\rho$, $\rho =
1.00 \pm 0.15$.  This is equivalent to $1.23 \le w_0 \le 1.66$, a range which
will be of particular interest to us.
\bigskip

\noindent
{\bf B.  Polarization and branching ratios} 
\medskip

\begin{figure} 
\begin{center} 
\includegraphics[height=7in]{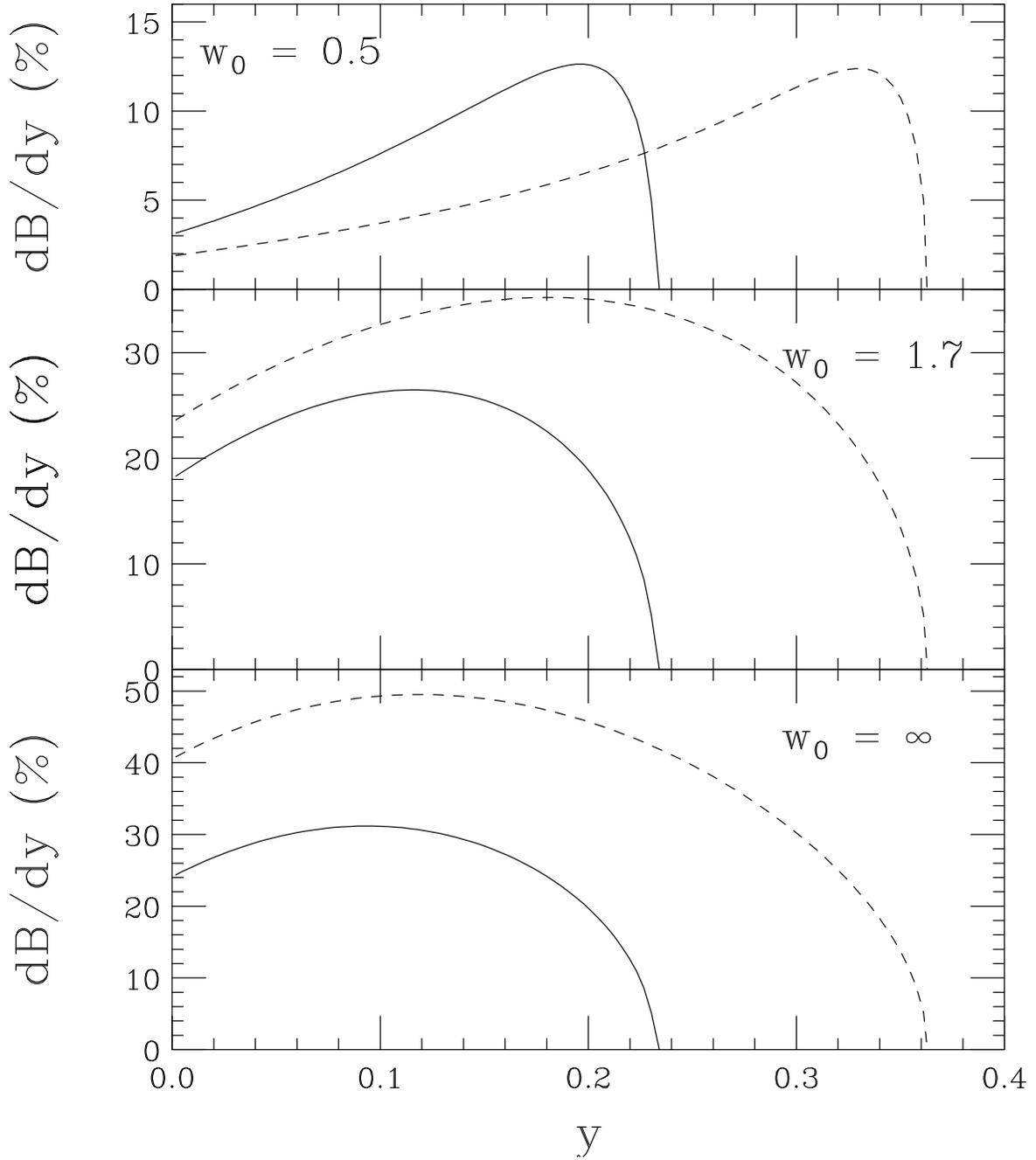} 
\end{center} 
\caption{Values of $dB/dy$ (in percent) for the processes $D_s^+\to \phi e^+ 
\nu_e$ (solid curves) and $D_s^+\to \omega_s e^+ \nu_e$ (dashed curves, where 
$\omega_s$ denotes a pure $s \bar s$ state with the mass of $\omega$), for 
several values of the form factor parameter $w_0$.  Top:  $w_0 = 0.5$; middle: 
$w_0 = 1.7$; bottom:  $w_0 = \infty$. 
\label{fig:ffall}} 
\end{figure} 

For the decays $D_s \to (\omega_s,\phi) \ell \nu$,
the sums $d\b/dy \equiv d\b_T/dy + d\b_L/dy$ are plotted for several values of 
$w_0$ in Fig.\ \ref{fig:ffall}.  The value $w_0 = 0.5$ is slightly below the
lowest value providing a fit to the total branching ratio $\b(D_s^+ \to \phi
e^+ \nu_e)$; the value $w_0 = 1.7$ is slightly above the highest value
considered for the universal form factor in Ref.\ \cite{Amundson:1992xp};
the value at $w_0 = \infty$ corresponds to no form factor damping. 
 
\begin{figure} 
\begin{center} 
\includegraphics[height=7in]{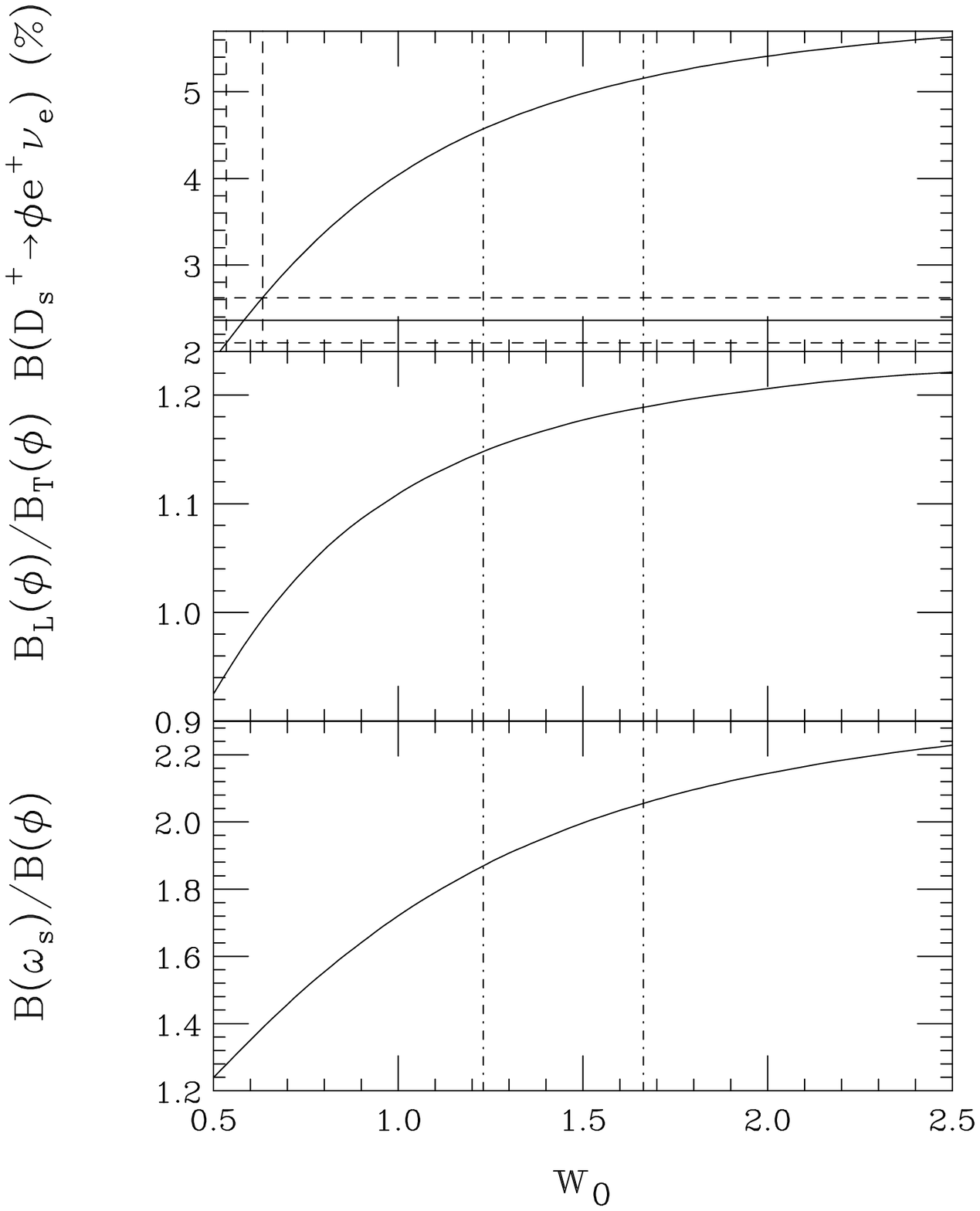} 
\end{center} 
\caption{Dependence on form factor parameter $w_0$ of various predicted 
quantities.  Vertical dash-dotted lines denote the limits on $w_0$ of the
universal form factor considered in Ref.\ \cite{Amundson:1992xp}.  Top:
$\b(D_s^+\to \phi e^+ \nu_e)$; middle:  ratio $\b_L(\phi)/ 
\b_T(\phi)$; bottom: $\b(\omega_s)/\b(\phi)$, where $\omega_s$ denotes a pure 
$s \bar s$ state with the mass of $\omega$.  In the top figure, the solid 
and dashed horizontal lines correspond to the central and $\pm 1 \sigma$ 
experimental values \cite{Amsler:2008zzb}, while the dashed vertical lines 
represent the corresponding $\pm 1 \sigma$ limits on $w_0$. 
\label{fig:br}} 
\end{figure}

The branching ratio $\b(D_s^+ \to \phi e^+ \nu_e) = \int_0^{y_{\rm max}} dy 
(d\b/dy)$, the ratio $\b_L(\phi)/ 
\b_T(\phi)$ of longitudinal to transverse decay rates, and the ratio $R_s =
\b(\omega_s)/\b(\phi)$ are plotted in Fig.\ \ref{fig:br}.  Values of $w_0$ 
between about 0.53 and 0.63 yield satisfactory values of $\b(D_s^+ \to \phi
e^+ \nu_e)$.  For $w_0 = (0.5,1.7,2.2,\infty)$ the ratios $R_s$ are
$(1.24,2.07,2.18,2.41)$.  We thus consider $1.2 \le R_s < 2.4$ 
as a conservative range. Although lower values are associated with values of
$w_0$ giving a better fit to $\b(D_s^+ \to \phi e^+ \nu_e)$, consideration of 
the semileptonic decays $D_s \to (\eta,\eta') \ell \nu$ and the hadronic
decay $D_s^+\to \phi \pi^+$ favors the higher ratio.

The ratio of longitudinal to transverse polarization in $D_s^+\to \phi e^+ 
\nu_e$ is predicted to range between 0.92 and 1.22 for $0.5 \le w_0 \le 2.5$, 
whereas the Particle Data Group average is quoted as $0.72 \pm 0.18$ 
\cite{Amsler:2008zzb}, based on the individual measurements
\cite{Kodama:1993,Frabetti:1994,Avery:1994} quoted in Table \ref{tab:LT}.
(Form factor ratios have been measured recently more precisely by the E791
\cite{Aitala:1998xu} and FOCUS \cite{Link:2004} Collaborations at Fermilab
and by the BaBar Collaboration at SLAC \cite{Aubert:2006ru},
but their ratios $\b_L(\phi)/\b_T(\phi)$ were not directly quoted.)
 
\begin{table}[h] 
\caption{Measurements of the ratio of longitudinal to transverse $\phi$ 
polarization in $D_s^+\to \phi \ell^+ \nu_\ell$. 
\label{tab:LT}} 
\begin{center} 
\begin{tabular}{c c c c} \hline \hline 
Reference & $\ell$ & Events & Ratio \\ \hline 
E653 \cite{Kodama:1993} & $\mu$ & 19 & $0.54 \pm 0.21 \pm 0.10$ \\ 
E687 \cite{Frabetti:1994} & $\mu$ & 90 & $1.0 \pm 0.5 \pm 0.1$ \\ 
CLEO \cite{Avery:1994} & $e$ & 308 & $1.0 \pm 0.3 \pm 0.2$ \\ 
Average \cite{Amsler:2008zzb} & & & $0.72 \pm 0.18$ \\ \hline \hline 
\end{tabular} 
\end{center} 
\end{table} 

Neglecting for a moment weak annihilation, the ratio $R \equiv \b(D_s^+ \to
\omega e^+ \nu_e)/\b(D_s^+ \to \phi e^+ \nu_e)$ is governed by several effects:
(1) a phase space correction, (2) a difference between form factors, and (3)
the $\omega-\phi$ mixing angle.  We have estimated that the product of the
first two gives a range $1.2 \le R_s \le 2.4$ for $\omega_s$ and $\phi_I$
composed entirely of $s \bar s$.  The mixing angle then implies $R = R_s \tan^2
\delta$, where $\tan^2 \delta = 3.41 \times 10^{-3}$ for $\delta=-3.34^\circ$.
We then find $R = (4.1-8.2)(\delta/3.34^\circ)^2 \times 10^{-3}$, implying
[when we take also $\pm 1 \sigma$ errors on $\b(D_s^+ \to \phi e^+ \nu_e)$]
that 
\beq 
\b(D_s^+ \to \omega e^+ \nu_e) = (0.9-2.1) \times 10^{-4} 
\left( \frac{\delta}{3.34^\circ} \right)^2~. 
\eeq 
While small, this branching ratio could be detectable in the present CLEO 
sample \cite{Alexander:2009ux} if backgrounds could be suitably suppressed 
and if $\delta$ were not anomalously small.

For completeness we discuss the decays $D_s \to (\eta,\eta') \ell \nu$.  In
principle these should be described by the same universal form factor as
$D_s \to \phi \ell \nu$, with $w^2$ in Eq.\ (\ref{eqn:wsq}) now defined as
\beq \label{eqn:wqsp}
w^2 = \frac{q^2 - q^2_{\rm max}}{M_{D_s} m_P} = \frac{q^2 - (M_{D_s} - m_P)^2}
{M_{D_s} m_P}~,
\eeq
where $P$ denotes the pseudoscalar meson ($\eta$ or $\eta'$).  The rather light
mass of the $\eta$ makes this approximation rather crude.  The assumption of a
universal pole in $w^2$ is not compatible with a universal pole in $q^2$, as
one sees from the definition of $w^2$.

\begin{figure}
\begin{center}
\includegraphics[width=0.7\textwidth]{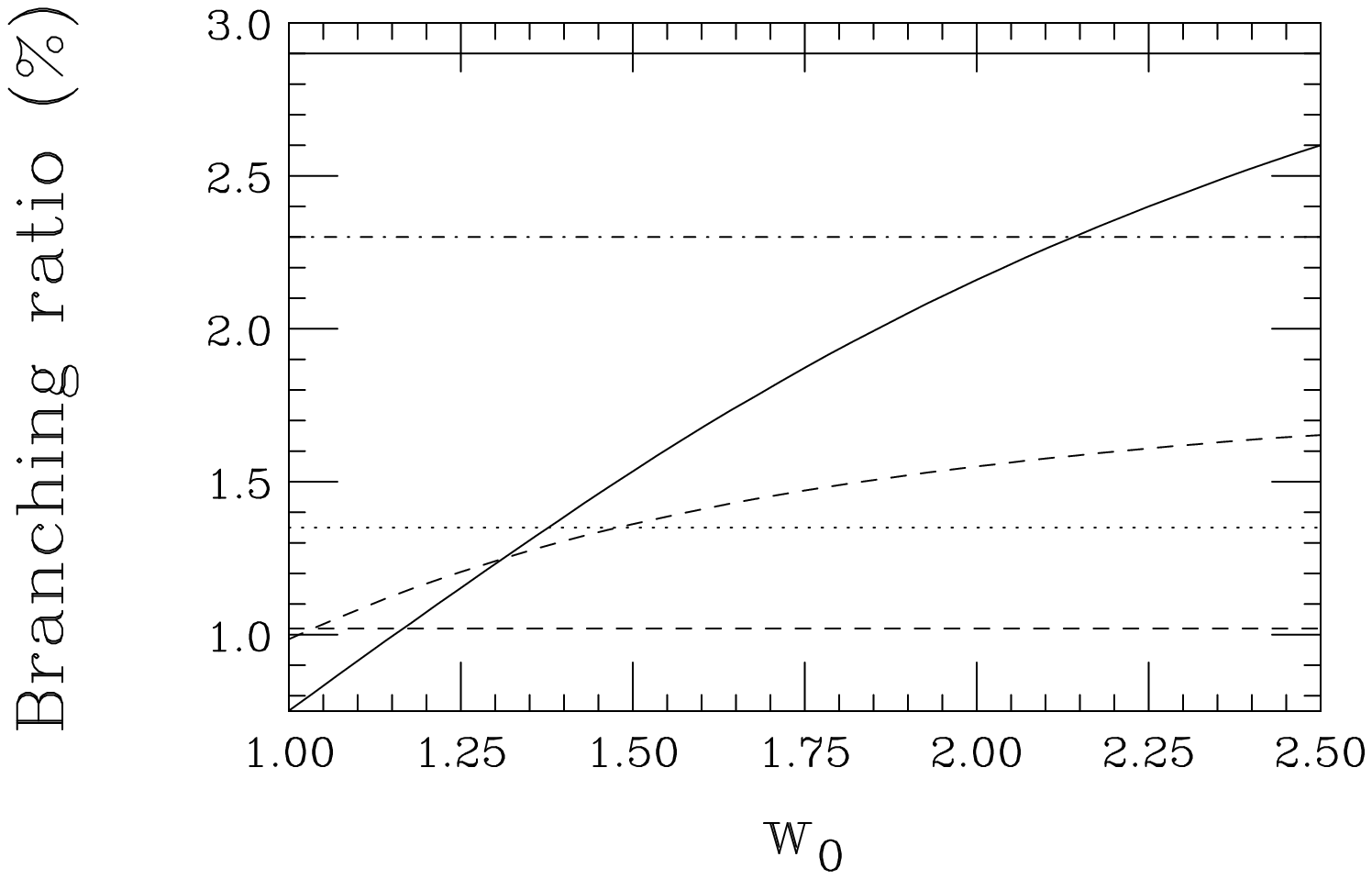}
\end{center}
\caption{Predicted branching ratios for $D_s \to \eta \ell \nu$ (solid curve)
and $(D_s \to \eta' \ell \nu$ (dashed curve) as a function of form factor
parameter $w_0$, with $\eta$ and $\eta'$ assigned the quark content
(\ref{eqn:mix}).  Horizontal solid and dashed lines denote central values
for $\b(D_s \to \eta \ell \nu)$ and $\b(D_s \to \eta' \ell \nu)$; horizontal
dotdashed and dotted lines denote, respectively, $- 1 \sigma$ and $+ 1 \sigma$
experimental limits for $\b(D_s \to \eta \ell \nu)$ and $\b(D_s \to \eta' \ell
\nu)$.
\label{fig:eta}}
\end{figure}

The observed branching ratios for $D_s$ semileptonic decays involving
$\eta$ and $\eta'$ are \cite{Amsler:2008zzb,Brandenburg:1995qq}
\beq
\b(D_s \to \eta \ell \nu) = (2.9 \pm 0.6)\%~,~~
\b(D_s \to \eta' \ell \nu) = (1.02 \pm 0.33)\%~.
\eeq
Charm nonleptonic decays \cite{Bhattacharya:2008ss,Bhattacharya:2008ke} and 
many other processes
involving $\eta$ and $\eta'$ are well-approximated by the mixing scheme
\beq \label{eqn:mix}
\eta \simeq \frac{1}{\st}(s \bar s - u \bar u - d \bar d)~,~~
\eta' \simeq \frac{1}{\sx} (2 s \bar s + u \bar u + d \bar d)~.
\eeq
With this scheme, the predicted branching ratios are plotted as functions
of $w_0$ in Fig.\ \ref{fig:eta}.

A successful fit to $\b(D_s \to \eta \ell \nu)$ at the $1 \sigma$ level
requires $w_0 > 2.1$, while a successful fit to $\b(D_s \to \eta' \ell \nu)$ at
the $1 \sigma$ level requires $w_0 < 1.5$.  This situation could be somewhat
improved if the mixing scheme (\ref{eqn:mix}) were altered so that
the strange quark admixture in the $\eta$ were increased while
the strange quark admixture in the $\eta'$ were decreased.  The scheme
(\ref{eqn:mix}) corresponds to an octet-single mixing angle of $\theta =
-\sin^{-1}(1/3) = -19.5^\circ$.  For the ISGW2 set of form factors
\cite{Scora:1995ty} considered in Ref.\ \cite{Brandenburg:1995qq}, $\theta
= -20^\circ$ leads to the prediction $\b(D_s^+ \to \eta' e^+ \nu_e)/
\b(D_s^+ \to \eta e^+ \nu_e) = 0.86$, to be compared with the measured value
of $0.35 \pm 0.09 \pm 0.07$.  Better agreement with the data is obtained
for $\theta = -10^\circ$, predicting this ratio to be 0.43.  This is very
close to the angle proposed by Isgur \cite{Isgurmix}, $\theta = -9.74^\circ$,
in which
\beq \label{eqn:Imix}
\eta \simeq \frac{1}{\s}s \bar s - \frac{1}{2}(u \bar u + d \bar d)~,~~
\eta' \simeq \frac{1}{\s}s \bar s + \frac{1}{2}(u \bar u + d \bar d)~.
\eeq
The $\eta'/\eta$ ratio in $D_s$ semileptonic decays for the scheme
(\ref{eqn:Imix}) is half that for (\ref{eqn:mix}).

\begin{figure}
\begin{center}
\includegraphics[width=0.7\textwidth]{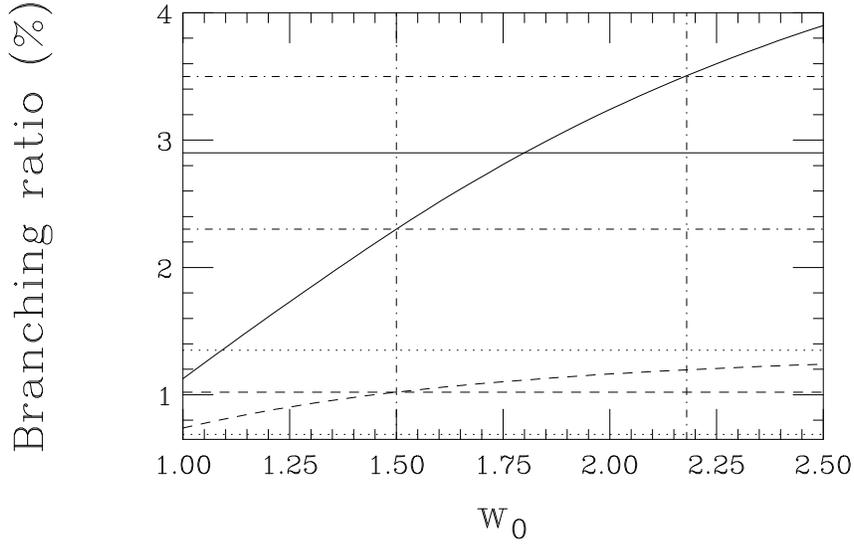}
\end{center}
\caption{Same as Fig.\ \ref{fig:eta} except that quark assignment
(\ref{eqn:Imix}) is used instead of (\ref{eqn:mix}).  Here horizontal
dotdashed and dotted lines denote, respectively, $\pm 1 \sigma$ and $\pm 1
\sigma$ experimental limits for $\b(D_s \to \eta \ell \nu)$ and $\b(D_s \to
\eta' \ell \nu)$.  Vertical dotdashed lines denote the limits $1.5 \le w_0
\le 2.18$ giving acceptable fits to both branching ratios.
\label{fig:etaI}}
\end{figure}

For the assignment (\ref{eqn:mix}), values of $w_0$ in the higher end of the
range 1.23--1.66 considered earlier seem to represent an acceptable compromise.
For $w_0=1.5$, the predicted $\eta'/\eta$ ratio is 0.89 for this scheme, while
for the assignment (\ref{eqn:Imix}), one predicts $\b(D_s^+ \to \eta e^+ \nu_e) =
2.30\%$, $\b(D_s^+ \to \eta' e^+ \nu_e) = 1.02$, with an $\eta'/\eta$ ratio of
0.44.  Values of $1.5 \le w_0 \le 2.18$ give acceptable fits to both
$\b(D_s^+ \to \eta e^+ \nu_e)$ and $\b(D_s^+ \to \eta' e^+ \nu_e)$ for the
assignment (\ref{eqn:Imix}), as illustrated in Fig.\ \ref{fig:etaI}.

\begin{figure}
\begin{center}
\includegraphics[width=0.6\textwidth]{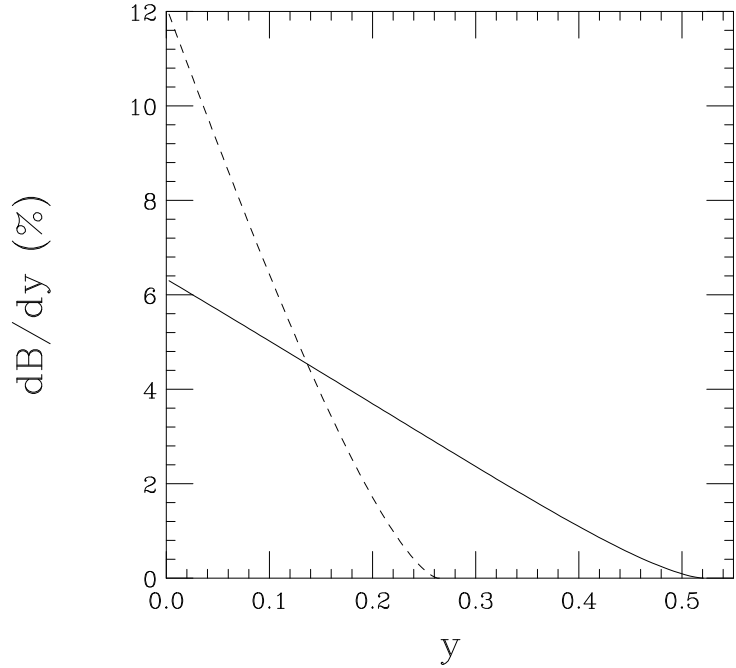}
\end{center}
\caption{Differential branching ratios (in percent) for $D_s \to \eta
\ell \nu$ (solid curve) and $D_s \to \eta' \ell \nu$ (dashed curve) for
$w_0 = 1.5$.  Here the assignment (\ref{eqn:mix}) has been used.  For the
assignment (\ref{eqn:Imix}), multiply the $\eta$ curve by 3/2 and the $\eta'$
curve by 3/4.
\label{fig:etaspec}}
\end{figure}

The spectra in $y =
q^2/M^2_{D_s}$ are compared for $D_s \to \eta \ell \nu$ and $D_s \to \eta'
\ell \nu$ in Fig.\ \ref{fig:etaspec} for $w_0=1.5$.  The enhancement of the
spectrum for $\eta'$ near $y=0$ with respect to that for $\eta$ represents the
(lesser, greater) recoil of the $(\eta',\eta)$ (and hence reflects a key aspect
of the heavy-quark theory), but may be exaggerated by the considerable
splitting of the $\eta$ and $\eta'$.
\bigskip

\centerline{\bf III.  RELATED HADRONIC PROCESSES} 
\bigskip 

\begin{figure}
\begin{center}
\includegraphics[width=0.8\textwidth]{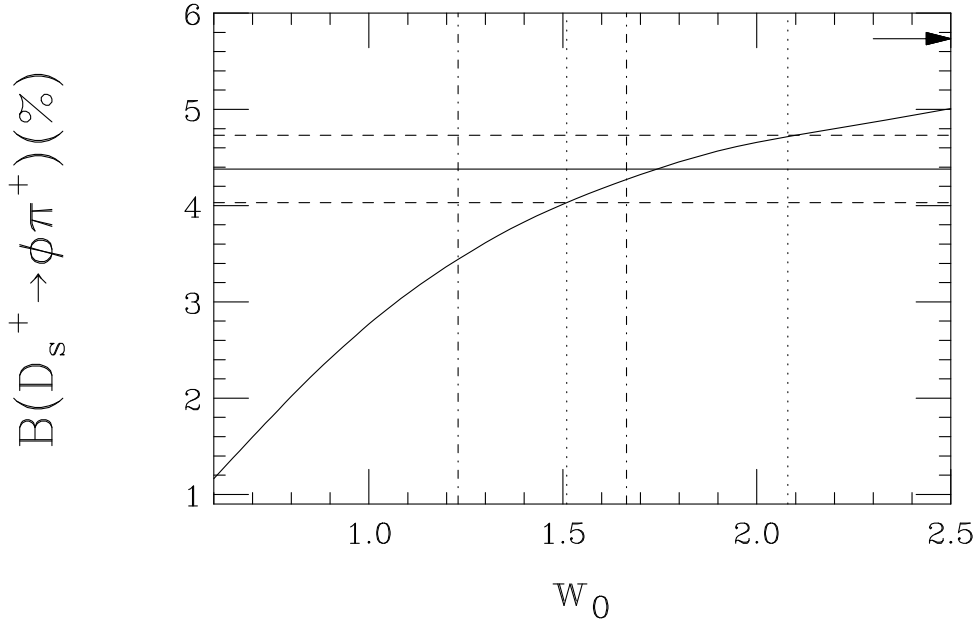}
\end{center}
\caption{Branching ratio for $D_s^+\to \phi \pi^+$ as a function of universal
form factor parameter $w_0$.  Horizontal solid and dashed lines denote central
and $\pm 1 \sigma$ experimental values \cite{Amsler:2008zzb}.  Vertical 
dash-dotted lines denote limits associated with universal monopole form factor
discussed in Ref.\ \cite{Amundson:1992xp},
while vertical dotted lines denote limits on $w_0$ based on $\pm 1 \sigma$
experimental values.  Arrow at upper right denotes
predicted branching ratio for $w_0 \to \infty$.
\label{fig:phipi}}
\end{figure}

We now consider $D_s^+\to \phi \pi^+$ in the heavy quark limit, again 
following Ref.\ \cite{Rosner:1990xx}.  The decay rate is predicted to be 
\beq 
\Gamma(D_s^+ \to \phi \pi^+) = \frac{[G_F V_{cs} V_{ud} f_\pi \xi(w^2_\pi) 
(1 + \sqrt{\zeta})]^2}{128 \pi \sqrt{\zeta}} M^3_{D_s} \lambda^{3/2}(1, 
\zeta,y_\pi)~, 
\eeq 
\beq 
w_\pi^2 \equiv \frac{m_\pi^2 - (M_{D_s} - m_\phi)^2}{M_{D_s}{m_\phi}} = 
-0.439~,~~y_\pi \equiv m_\pi^2/M^2_{D_s} = 5.03 \times 10^{-3}~. 
\eeq
Using $V_{cs} = V_{ud} = 0.974$, $f_\pi = 130.4$ MeV~\cite{RS}, we find 
\beq
\b(D_s^+\to \phi \pi^+) = \frac{5.73\%}{(1 - w_\pi^2/w_0^2)^2}~. 
\eeq
We plot this quantity as a function of $w_0$ in Fig.\ \ref{fig:phipi}.

As the experimental branching ratio is~\cite{Amsler:2008zzb} 
\beq \label{eqn:PDGBR}
\b(D^+_s\to\phi\pi^+) = (4.38 \pm 0.35)\%~,
\eeq 
only a modest form factor suppression can be tolerated, 
whereas the value of $w_0 \simeq 0.6$ leading to an acceptable branching ratio 
for $D_s^+\to \phi e^+ \nu_e$ implies $\b(D_s^+\to \phi \pi^+) = 1.2\%$.  We 
are thus led to consider the conservative limits $0.5 \le w_0 \le \infty$ 
in obtaining the range $1.2 \le R_s \le 2.4$ mentioned above.
If we were to allow a fit to $\b(D_s^+\to \phi \pi^+)$ at the $\pm 1 \sigma$
level while demanding better agreement with other decays, we could demand
$w_0 < 2.1$ (see Fig.\ \ref{fig:phipi}).  This would only reduce the
upper limit on $R_s$ by about 10\%.

It has been argued that the $K^+ K^-$ S-wave contribution in $D_s^+ \to K^+ K^-
\pi^+$ cannot be overlooked \cite{Stone:2008ak}, with
\beq
\frac{\Gamma(D_s^+ \to f_0(980) \pi^+ \to K^+ K^- \pi^+)}
     {\Gamma(D_s^+ \to \phi \pi^+ \to K^+ K^- \pi^+)} = 0.3 \pm 0.1~.
\eeq
Applying this correction to the branching ratio (\ref{eqn:PDGBR}), one obtains
$\b(D^+_s\to\phi\pi^+) = (4.38 \pm 0.35)\%/(1.3 \pm 0.1) = (3.37 \pm 0.37)\%$,
implying $1.07 \le w_0 \le 1.36$, still within our range of consideration.
 
The hadronic process $D_s^+ \to  \omega \pi^+$ would be related to $D_s^+ \to 
\omega e^+ \nu_e$ if the only contributing amplitude were the color-favored 
subprocess $c \to s \pi^+$ followed by the mixing transition $s \bar s \to 
(u \bar u + d \bar d)/\s$ giving an $\omega$ in the final state.  This is not 
the case, however.  A factorization calculation based on this assumption would 
predict 
\beq 
\frac{\b(D_s^+ \to \omega \pi^+)}{\b(D_s^+ \to \phi \pi^+)} = \left( 
\frac{p^*_{\pi \omega}}{p^*_{\pi \phi}} \right)^3 \tan^2 \delta~, 
\eeq 
where $p^*_{\pi \omega} = 822$ MeV/$c$ and $p^*_{\pi \phi} = 712$ MeV/$c$ are 
center-of-mass 3-momenta for the respective decays.
With $\b(D_s^+ \to \phi \pi^+) = 
(4.38 \pm 0.35)\%$, this implies $\b(D_s^+ \to \omega \pi^+) = 
(2.3 \pm 0.2) \times 10^{-4}(\delta/3.34^\circ)^2$.  The experimental value 
is considerably larger \cite{Amsler:2008zzb,Balest:1997pp}, 
\beq
\b(D_s^+ \to \omega \pi^+) = (2.5 \pm 0.9) \times 10^{-3}~,
\eeq
implying the importance of a weak annihilation contribution
\cite{Bhattacharya:2008ke}. 

As has been mentioned, the $c \bar s \to u \bar d$ ``annihilation'' amplitude
$A$, if interpreted literally, would be subject to helicity suppression, so in
flavor SU(3) treatments \cite{Bhattacharya:2008ss,Bhattacharya:2008ke} it
must represent a shorthand for rescattering contributions.  Further evidence
for this viewpoint comes from the observation of the decay $D_s^+ \to p \bar n$
\cite{Athar:2008ug}.  If interpreted literally in terms of the production
of $p \bar n$ by the weak current from $c \bar s$ annihilation (i.e., if
treated by a factorization hypothesis), this process would be highly
suppressed by PCAC \cite{Pham:1980}, whereas the observed branching ratio is
$\b(D_s^+ \to p \bar n) = (1.30 \pm 0.36^{+0.12}_{-0.16}) \times 10^{-3}$
\cite{Athar:2008ug}.  

The decays $D_s^+ \to \eta \pi^+$ and $D_s^+ \to \eta' \pi^+$ may be related to
$D_s^+ \to \phi \pi^+$ in the heavy-quark limit.  
(For a study of $D_s^+\to \eta\pi^+$ and $D_s^+ \to \eta'\pi^+$ using
factorization of the tree amplitude, see Ref.\ \cite{Kamal:1988cd}.)
In the treatment of Ref.\ \cite{Rosner:1990xx}, 
the ratio of partial widths contributed by the factorized
tree (``$T$'') amplitude is given in the limit of degenerate $s \bar s$ vector
$V_s$ and pseudoscalar $P_s$ masses by
\beq
\frac{\Gamma(D_s^+ \to V_s \pi^+)_T}{\Gamma(D_s^+ \to P_s \pi^+)_T}
= \left[ \frac{1 + \sqrt{\zeta}}{1 - \sqrt{\zeta}} \right]^2
\frac{\lambda(1, \zeta, y_\pi)}{[(1 + \sqrt{\zeta})^2 - y_\pi^2]^2}~,
\eeq
where $\sqrt{\zeta} \equiv M_{V,P}/M_{D_s}$ and $y_\pi \equiv m^2_\pi/
M^2_{D_s}$.  Neglecting the small quantity $y_\pi$, we find in this limit
that the right-hand side reduces to unity, so
\beq
\Gamma(D_s \to V_s \pi^+)_T = \Gamma(D_s \to P_s \pi^+)_T~,
\eeq
or, independently of the precise nature of octet-singlet mixing in $\eta$ and
$\eta'$, and neglecting phase space differences,
\beq \label{eqn:sr}
\b(D_s^+ \to \phi \pi^+)_{T} = \b(D_s^+ \to \eta \pi^+)_T
  + \b(D_s^+ \to \eta' \pi^+)_T~.
\eeq

The decay $D_s^+ \to \phi \pi^+$ is expected to be dominated by the $T$
amplitude \cite{Bhattacharya:2008ke}, while small corrections to $T$
dominance are due to the annihilation amplitude $A$ in $D_s^+ \to (\eta,\eta')
\pi^+$ \cite{Bhattacharya:2008ss}.  The branching ratios for $D_s^+ \to (\eta,
\eta') \pi^+$ are \cite{Amsler:2008zzb}
\beq
\b(D_s^+ \to \eta \pi^+) = (1.58 \pm 0.21)\%~,~~
\b(D_s^+ \to \eta' \pi^+) = (3.8 \pm 0.4)\%~,
\eeq
while the contributions of the tree amplitudes to these decay widths are
\cite{Bhattacharya:2008ss}
\bea
\b(D_s^+ \to \eta \pi^+)_T & = & \left( \frac{1.605}{1.50} \right)^2
(1.58 \pm 0.21)\% = (1.81 \pm 0.24)\%~, \cr
\b(D_s^+ \to \eta' \pi^+)_T & = & \left( \frac{2.27}{2.55} \right)^2
(3.8 \pm 0.4)\% = (3.01 \pm 0.32)\%.
\eea
The sum rule (\ref{eqn:sr}) then reads
\beq
(4.38 \pm 0.35)\% = (4.82 \pm 0.40)\%~,
\eeq
which is satisfactorily obeyed.  A similar confirmation of the heavy-quark
relation between tree amplitudes in $PP$ and $VP$ decays of charmed mesons
was obtained in Refs.\ \cite{Bhattacharya:2008ss} 
and \cite{Bhattacharya:2008ke} by comparing their
contributions in $D \to \overline{K} \pi$ and $D \to \overline{K^*}\pi$ decays.
(See, in particular, Eqs.\ (18) and (19) in Ref.\ \cite{Bhattacharya:2008ke}.)
\bigskip 

\centerline{\bf IV.  WEAK ANNIHILATION IN $D_s^+ \to \omega \pi^+$ and $D_s^+ \to
\omega e^+ \nu_e$.}
\bigskip

A difficulty (see, e.g., Refs.\ \cite{Bhattacharya:2008ke,Rosner:1999xd,Cheng})
in ascribing the decay $D_s^+ \to \omega \pi^+$ to the weak subprocess $c\bar s
\to u \bar d$ is that because the final $u \bar d$ state has odd G-parity (as
does a pion), it cannot decay to $\omega \pi^+$, which has even G-parity
\cite{Lipkin:1989vp,Kamal:1988wj,Fajfer:2003}.  In a flavor-symmetric
description \cite{Rosner:1999xd},
the decays $D_s^+ \to \rho^0 \pi^+$ and $D_s^+ \to \omega \pi^+$ both involve
amplitudes $A_V$ and $A_P$, where the subscript denotes whether the $\bar d$
quark in the $c \bar s \to u \bar d$ subprocess is included in a pseudoscalar
($P$) or a vector ($V$) meson.  These are required to cancel one another
for $D_s^+ \to \omega \pi^+$ in order to enforce the G-parity selection rule;
they will then add in $D_s^+ \to \rho^0 \pi^+$.  However, one sees a branching
ratio $\b(D_s^+ \to \omega \pi^+) = (2.5 \pm 0.9) \times 10^{-3}$, while 
$D_s^+ \to \rho^0 \pi^+$ is only quoted as ``not seen'' \cite{Amsler:2008zzb}.

Moreover, annihilation topologies, if interpreted literally in terms of
quarks, are subject to helicity selection rules leading to their suppression,
so one must interpret them as encoding the effects of rescattering.  The
authors of Ref.\ \cite{Kamal:1988wj} ascribe the decay $D_s^+ \to \omega \pi^+$
to the weak decay $D_s^+ \to \overline{K}^{(*)0} K^{(*)+}$ followed by
$\overline{K}^{(*)0} K^{(*)+} \to \omega \pi^+$.  A successful prediction of
the branching ratio for $D_s^+ \to \omega \pi^+$ was made on the basis of
final-state interactions in Ref.\ \cite{Buccella}.  However, 
within these two frameworks there is no
corresponding process contributing to $D_s^+ \to \omega \ell^+ \nu$.

An alternate possibility is that the decay $D_s^+ \to \omega \pi^+$ proceeds
through pre-radiation of the $\omega$, whether via violation of the OZI
rule or rescattering.  An example of the latter mechanism would be the
dissociation of the $D_s^+$ into two-meson states such as $D^{(*)0} K^{(*)+}$
and $D^{(*)+} K^{(*)0}$.  The two mesons can be $PV$, $VP$, or $VV$ and must be
in a relative P-wave; $PP$ is forbidden by parity.  The two mesons then
rescatter strongly to $(c \bar s) \omega$ and the virtual $c \bar s$ state
decays weakly to $\pi^+$.

A correponding mechanism can generate the decay $D_s^+ \to \omega e^+ \nu_e$.
Here, the virtual $c \bar s$ (which can now be spin-1, and hence not subject to
helicity suppression) decays to a lepton pair.  We may estimate
very crudely the branching ratio for this process if the corresponding
process (described above) is responsible for $D_s^+ \to \omega \pi^+$.
Neglecting all kinematic factors, we expect
\beq
\frac{\b(D_s \to \omega \ell \nu)}{\b(D_s \to \phi \ell \nu)} =
\frac{\b(D_s^+ \to \omega \pi^+)}{\b(D_s^+ \to \phi \pi^+)}~.
\eeq
Using the branching ratios quoted earlier, we infer
\beq
\b(D_s \to \omega \ell \nu)_{\rm WA} = (1.3 \pm 0.5) \times 10^{-3}~,
\eeq
roughly an order of magnitude larger than one would conclude if
$\omega$--$\phi$ mixing were solely responsible for the decay.

We have neglected differences in form factor behavior which are to be
expected for the WA process, since it is expected to be peaked at
maximum $q^2$.  This peaking occurs both in the scenario where the
$\omega$ is emitted via an OZI-suppressed three-gluon coupling from
the initial $c \bar s$ system, and where rescattering gives rise to a
virtual $D_s^*$ which then decays to $\ell \nu$.  In the latter case,
high $q^2$ is favored by proximity to the $D_s^*$ pole.
By contrast, as can be seen in Fig.\ \ref{fig:ffall},
one does not expect peaking for $D_s \to \phi \ell \nu$ at high $q^2$
except for the lowest values of $w_0$.  The peaking of the spectrum for
$D_s \to \omega \ell \nu$ at maximum $q^2$ will be one of the hallmarks of
the WA process.
\bigskip

\centerline{\bf V.  CONCLUSION} 
\bigskip 
 
We have considered the ratio $R = \b(D_s^+ \to \omega e^+ \nu_e)/\b(D_s^+ \to 
\phi e^+ \nu_e)$ as a test of $\phi$--$\omega$ mixing in the absence of 
nonperturbative enhancements, and, in the event that the ratio exceeds a 
nominal estimate, as possible evidence for such enhancements, termed 
``weak annihilation'' \cite{WA}.  We find for $\phi$--$\omega$ mixing a range
\beq 
R = (4.1-8.2)(\delta/3.34^\circ)^2 \times 10^{-3}~, 
\eeq 
where $\delta$ is the $\omega$--$\phi$ mixing angle.  The value $\delta = 
3.34^\circ$ is obtained in one mass-independent analysis 
\cite{Benayoun:1999fv}, while a considerably smaller value of $-0.45^\circ$ at 
$m_\omega$ is found when the angle is allowed to vary with mass 
\cite{Benayoun:2007cu}.  

Given the experimental branching ratio 
$\b(D_s^+ \to \phi e^+ \nu_e) = (2.36 \pm 0.26)\%$, we conclude that any 
value of $\b(D_s^+ \to \omega e^+ \nu_e)$ exceeding $(8.2 \times 10^{-3}) 
\cdot (2.6\%) \simeq 2 \times 10^{-4}$ is unlikely to be explainable via 
$\omega$--$\phi$ mixing, and would provide evidence for nonperturbative 
effects such as those discussed in Refs.\ \cite{WA}.  A crude estimate
based on comparing hadronic and semileptonic processes gives a branching
ratio $\b(D_s^+ \to \omega e^+ \nu_e) = (1.3 \pm 0.5)\times 10^{-3}$, nearly
an order of magnitude higher than the values from $\omega$--$\phi$ mixing
alone.
\bigskip 
 
\centerline{\bf ACKNOWLEDGMENTS} 
\bigskip 

We thank I. Bigi and J. Wiss for helpful discussions. 
This work was supported in part by 
the United States Department of Energy through Grant No.\ DE-FG02-90ER-40560. 


\bigskip
 
\begin{thebibliography}{99} 
 
\bibitem{Alexander:2009ux} 
See, e.g., J.~P.~Alexander {\it et al.} [CLEO Collaboration], 
  arXiv:0901.1216 [hep-ex]. 
 
\bibitem{Amsler:2008zzb} 
  C.~Amsler {\it et al.}  [Particle Data Group], 
  Phys.\ Lett.\  B {\bf 667}, 1 (2008). 
 
\bibitem{Okubo:1963fa} S.~Okubo, 
Phys.\ Lett.\ {\bf 5}, 165 (1963); 
G. Zweig, CERN Report No. 8419/TH--412 (1964); 
J.~Iizuka, Prog.\ Theor.\ Phys.\ Suppl.\ {\bf 37}, 21 (1966). 

\bibitem{WA} 
I. I. Y. Bigi, Z. Phys.\ C {\bf 5} 313 (1979); Z. Phys.\ C {\bf 9}, 197 (1981);
 Nucl.\ Phys.\ {\bf B177}, 395 (1981);
I. I. Y. Bigi and M. Fukugita, Phys.\ Lett.\ {\bf 97B}, 121 (1980);
I. I. Bigi and N. G. Uraltsev, Phys.\ Lett.\ B {\bf 280}, 271 (1992);
 Nucl.\ Phys.\ {\bf B423}, 33 (1994);
M. Neubert and C. T. Sachrajda, Nucl.\ Phys.\ {\bf B483}, 339 (1997); 
M. B. Voloshin, Phys.\ Lett.\ B {\bf 515}, 74 (2001);
A. K. Leibovich, Z. Ligeti, and M. B. Wise, Phys.\ Lett.\ B {\bf 539}, 242
 (2002); 
H.-Y. Cheng, Eur.\ Phys.\ J.\ C {\bf 26}, 551 (2003); 
S. Fajfer, A. Prapotnik, P. Singer, and J. Zupan, Phys.\ Rev.\ D {\bf 68},
094012 (2003); 
P. Gambino, J. Ossola, and N. Uraltsev, JHEP {\bf 0509}, 010 (2005);
P.~Gambino, P.~Giordano, G.~Ossola and N.~Uraltsev,
JHEP {\bf 0710}, 058 (2007). 
Further references may be found in
S.~Bianco, F.~L.~Fabbri, D.~Benson and I.~Bigi,
  Riv.\ Nuovo Cim.\ {\bf 26N7}, 1 (2003).

\bibitem{CLEOWA} J. L. Rosner {\it et al.} (CLEO Collaboration), Phys.\ Rev.\ 
Lett.\ {\bf 96}, 121801 (2006). 
 
\bibitem{Aubert:2007tw} 
B.~Aubert {\it et al.}  [BABAR Collaboration], 
  arXiv:0708.1753 [hep-ex]. 

\bibitem{Lipkin:1989vp}
  H.~J.~Lipkin,
  AIP Conf.\ Proc.\ {\bf 196}, 72 (1989).  See also:
  H. J. Lipkin, Phys.\ Lett.\ B {\bf 283}, 412 (1992), and {\it Proceedings
  of the 2nd International Conference on $B$ Physics and CP Violation,}
  Honolulu, Hawaii, edited by T. E. Browder {\it et al.} (World Scientific,
  Singapore, 1998), p.\ 436; H.-Y. Cheng, Ref.\ \cite{WA}.

\bibitem{Gronau:2008kk} M.~Gronau and J.~L.~Rosner, 
  Phys.\ Lett.\  B {\bf 666}, 185 (2008). 
 
\bibitem{Benayoun:1999fv} 
M.~Benayoun, L.~DelBuono, S.~Eidelman, V.~N.~Ivanchenko and H.~B.~O'Connell, 
Phys.\ Rev.\ D {\bf 59}, 114027 (1999).
See also A.~Kucukarslan and U.~G.~Meissner, 
Mod.\ Phys.\ Lett.\ A {\bf 21}, 1423 (2006).
 
\bibitem{Benayoun:2007cu} 
M.~Benayoun, P.~David, L.~DelBuono, O.~Leitner and H.~B.~O'Connell, 
Eur.\ Phys.\ J.\  C {\bf 55}, 199 (2008).

\bibitem{HQS} The literature on this subject is vast.  A partial list:
S. Nussinov and W. Wetzel, Phys.\ Rev.\ D {\bf 36}, 130 (1987); M. B. Voloshin
and M. A. Shifman, Yad.\ Phys.\ {\bf 45}, 463 (1987) [Sov.\ J. Nucl.\ Phys.\
{\bf 45}, 292 (1987)]; Yad.\ Phys.\ {\bf 47}, 801 (1988) [Sov.\ J. Nucl.\
Phys.\ {\bf 47}, 511 (1988)]; N. Isgur, D. Scora, B. Grinstein, and M. B. Wise,
Phys.\ Rev.\ D {\bf 39}, 799 (1989); N. Isgur and M. B. Wise, Phys.\ Lett.\
B {\bf 232}, 113 (1989); {\bf 237}, 527 (1990); H. Georgi, Phys.\ Lett.\
{\bf 240}, 447 (1990); A. F. Falk, H. Georgi, B. Grinstein, and M. B. Wise,
Nucl.\ Phys.\ {\bf B343}, 1 (1990); J. D. Bjorken, Report No.\ SLAC-PUB-5278
(unpublished); invited talk presented at Les Rencontres de Physique de la
Valle d'Aoste, La Thuile, Aosta Valley, Italy, 1990 (unpublished).  Our
parametrization of form factors is based on this last work.

\bibitem{Rosner:1990xx} J.~L.~Rosner, Phys.\ Rev.\ D {\bf 42}, 3732 (1990). 

\bibitem{Amundson:1992xp}
  J.~F.~Amundson and J.~L.~Rosner,
  Phys.\ Rev.\  D {\bf 47}, 1951 (1993).
 
\bibitem{QCDcorr} M. B. Voloshin and M. A. Shifman, Yad.\ Fiz.\ {\bf 45}, 
463 (1987) [Sov.\ J.\ Nucl.\ Phys.\ {\bf 45}, 292 (1987)]; H. D. Politzer and 
M. B. Wise, Phys.\ Lett.\ B {\bf 206}, 681 (1988); {\bf 208}, 504 (1988). 

\bibitem{Kodama:1993} K. Kodama {\it et al.} (Fermilab E653 Collaboration), 
Phys.\ Lett.\ B {\bf 309}, 483 (1993). 
 
\bibitem{Frabetti:1994} P. L. Frabetti {\it et al.} (Fermilab E687 
Collaboration), Phys.\ Lett.\ B {\bf 328}, 187 (1994). 
 
\bibitem{Avery:1994} P. Avery {\it et al.} (CLEO Collaboration), Phys.\ Lett.\ 
B {\bf 337}, 405 (1994). 

\bibitem{Aitala:1998xu}
  E.~M.~Aitala {\it et al.}  [E791 Collaboration],
  Phys.\ Lett.\  B {\bf 450}, 294 (1999).

\bibitem{Link:2004} J. M. Link {\it et al.} (FOCUS Collaboration), Phys.\
Lett.\ B {\bf 586}, 183 (2004).

\bibitem{Aubert:2006ru}
B.~Aubert {\it et al.} [BABAR Collaboration],
Stanford Linear Accelerator Center Report SLAC-PUB-12017,
  arXiv:hep-ex/0607085, contributed to 33rd International Conference on High
  Energy Physics (ICHEP 06), Moscow, Russia, 2006.

\bibitem{Brandenburg:1995qq}
  G.~Brandenburg {\it et al.}  [CLEO Collaboration],
  Phys.\ Rev.\ Lett.\  {\bf 75}, 3804 (1995).
 
  \bibitem{Bhattacharya:2008ss}
  B.~Bhattacharya and J.~L.~Rosner,
  Phys.\ Rev.\  D {\bf 77}, 114020 (2008).
  
  \bibitem{Bhattacharya:2008ke}
B.~Bhattacharya and J.~L.~Rosner, arXiv:0812.3167 [hep-ph], to be published
in Phys.\ Rev.\ D. 

\bibitem{Scora:1995ty}
  D.~Scora and N.~Isgur,
  Phys.\ Rev.\ D {\bf 52}, 2783 (1995).

\bibitem{Isgurmix}
N. Isgur, Phys.\ Rev.\ D {\bf 13}, 122 (1976).
 
  \bibitem{RS} J. L. Rosner and S. Stone, review of meson decay constants 
in \cite{Amsler:2008zzb}. 

\bibitem{Stone:2008ak}
  S.~Stone and L.~Zhang,
  arXiv:0812.2832 [hep-ph].
 
\bibitem{Balest:1997pp}
  R.~Balest {\it et al.}  [CLEO Collaboration],
  Phys.\ Rev.\ Lett.\  {\bf 79}, 1436 (1997).
  
  \bibitem{Athar:2008ug}
  S.~B.~Athar {\it et al.}  [CLEO Collaboration],
  Phys.\ Rev.\ Lett.\  {\bf 100}, 181802 (2008).

\bibitem{Pham:1980} X. Y. Pham, Phys.\ Rev.\ Lett.\ {\bf 45}, 1663 (1980).
 
\bibitem{Kamal:1988cd}
  A.~N.~Kamal, N.~Sinha and R.~Sinha,
  Phys.\ Rev.\  D {\bf 38}, 1612 (1988).
 
\bibitem{Rosner:1999xd}
J.~L.~Rosner,
  Phys.\ Rev.\  D {\bf 60}, 114026 (1999).
  
  \bibitem{Cheng}
  H.~Y.~Cheng, Ref.\ \cite{WA}

\bibitem{Kamal:1988wj}
A.~N.~Kamal, N.~Sinha and R.~Sinha,
 Phys.\ Rev.\  D {\bf 39}, 3503 (1989).

\bibitem{Fajfer:2003} S. Fajfer {\it et al.}, Ref.\ \cite{WA}. 

\bibitem{Buccella} F. Buccella, M. Lusignoli, and A. Pugliese, 
Phys.\ Lett.\ B {\bf 379}, 249 (1996).

\end{thebibliography}
\end{document}